\documentstyle[preprint,pra,aps,epsfig]{revtex}
\begin{document}
\tighten
\def\question#1{{{\marginpar{\tiny \sc #1}}}}
\draft
\title{NewtonPlus: Approximate Relativity for Supernova Simulations}
\author{Christian Y. Cardall$^{1,2,3}$, Anthony Mezzacappa$^1$,
and Matthias Liebend\"orfer$^{1,2}$}
\address{$^1$Physics Division, Oak Ridge National Laboratory, Oak Ridge,
 	TN 37831-6354 \\
	$^2$Department of Physics and Astronomy, University of Tennessee,
	Knoxville, TN 37996-1200 \\ 
	$^3$Joint Institute for Heavy Ion Research, Oak Ridge National
	Laboratory, Oak Ridge, TN 37831-6374}
\date{June 4, 2001}
\maketitle

\begin{abstract}
We propose an approximation to general relativity that captures the 
main gravitational effects of dynamical importance in 
supernovae. The
conceptual link between this formalism and the Newtonian limit is
such that it could likely be implemented 
in existing
multidimensional
Newtonian gravitational hydrodynamics codes employing a Poisson solver.
As a test of the formalism's utility, we display results for rapidly rotating
(and therefore highly deformed) neutron stars.
\end{abstract}


\section{Introduction}
\label{sec:intro}

The collapsed cores of massive stars are relativistic bodies.
In addition to relativistic effects, 
convection and rapid rotation may play important roles in core
collapse supernovae, making their accurate simulation a 
three-dimensional (3D) endeavor. Because accurate and stable 
numerical solutions to 3D relativistic problems 
are difficult to achieve, an approximation that captures the relativistic
phenomena most relevant to supernovae is desirable.
(More detailed discussion of these points, together with references, 
are given in the following
paragraphs.)

While relativistic treatments would be necessary to follow
the detailed processes leading to black hole formation, even collapse
events leading to neutron stars ought to be treated relativistically,
particularly at late times.
Denoting the gravitational potential by $\Phi$,
the Newtonian limit is obtained from the Einstein equations using the
metric 
\begin{equation}
   ds^2 = -(1+2\Phi) dt^2+(1-2\Phi) d{\bf x}^2,
\label{newtonMetric}
\end{equation} 
provided
$\Phi\ll 1$ and velocities are much less than the speed of light.
Taking $M$ and $R$ to be the mass and radius of the collapsed core, 
the gravitational potential is expected to be of order
\begin{equation}
\Phi\sim- {M\over R}=-0.21\left(M\over 1.4 \ M_\odot\right)
	\left(R\over 10{\rm\ km}\right)^{-1}, 
\label{potentialEstimate}
\end{equation}
while infall and/or ejecta velocities of order
\begin{equation}
v_{\rm ejecta}\sim \sqrt{2M\over R}=0.45\left(M\over 1.4 \ M_\odot\right)^{1/2}
	\left(R\over 10{\rm\ km}\right)^{-1/2}
\end{equation}
might be encountered. If some collapsed cores are born with rotation periods
comparable to observed millisecond pulsars, equatorial velocities 
would be of order
\begin{equation}
v_{\rm equator}=\Omega R = 0.21\left(T\over {\rm ms}\right)^{-1}
	\left(R\over 10{\rm\ km}\right).
\end{equation}
Crudely taking the core as a cold degenerate gas of nucleons of mass $m_B$
and Fermi momentum $p_F$, nucleon velocities of order
\begin{equation}
v_{\rm nucleon}\sim {p_F\over m_B}=0.69 \left(M\over 1.4 \ M_\odot\right)^{1/3}
	\left(R\over 10{\rm\ km}\right)^{-1}
\label{nucleonVelocity}
\end{equation}
clearly indicate that pressure and internal energy density 
cannot be neglected as gravitational sources.
The numerical coefficients in equations 
(\ref{potentialEstimate}-\ref{nucleonVelocity}) may seem to overstate the
case for relativity immediately after core bounce, when a less massive inner
core has a larger radius. They are perhaps more relevant 
numbers for late times,
when the final explosion energy \cite{burr95} and remnant nature (neutron star
vs. black hole) are determined.  
However, even at earlier times relativity cannot
be ignored: detailed comparisons of 
Newtonian and relativistic simulations in spherical symmetry show
more compact cores and higher neutrino luminosities and average energies
in relativistic treatments \cite{brue01,lieb00}. 
Because of indications that small (several percent) 
variations in, for example, neutrino
luminosities and shock stagnation radius 
can make the difference between a successful (exploding) and a failed
supernova 
\cite{jank96}, it is clear
that even modest relativistic effects 
comprise an indispensible component of realism in supernova studies.

The multidimensional nature of supernovae must also be recognized as
an important aspect of realism. Various observations, especially
data from SN 1987A, point to the asphericity of supernova explosions
(see e.g. \cite{jank96} for an overview, and \cite{wang01} for a
recent polarimetry analysis of several supernovae). 
Convection---either deep in the core 
and driven by a lepton gradient \cite{epst79}
or doubly-diffusive phenomena (``neutron fingers'') \cite{smar81},
or in the more tenuous region between the 
neutrinosphere and the stalled shock, driven 
by an entropy gradient \cite{arne87,beth87}---has for some time
been suggested as a means of boosting neutrino luminosities or increasing
the efficiency of heating the material behind the stalled shock.
The results of various simulations in up to two dimensions (2D)
differ as to the significance
of convective effects 
\cite{wils93,mill93,hera94,burr95,jank96,mezz98a,mezz98b}; 
clearly further study is necessary. Beyond convection, 
a pioneering 2D simulation studied jet production by magnetic fields 
in a supernova context \cite{lebl70}, and simulations of explosions from
collapsed stars with phenomenologically introduced jets have also been
performed in 2D \cite{macf99,khok99,macf01}.

These 2D simulations have yielded interesting results,
but ultimately consideration of the third dimension will be necessary.
Based on an initial exploration of 3D effects, it has been reported 
\cite{jank96}
that the sizes of convective cells in 3D simulations are about half as large
as in 2D simulations. Moreover, rapid rotation can significantly affect
the strength and spatial distribution of convection \cite{frye00}.
Detailed studies of magnetic field generation, 
jet formation, and neutron star kicks also invite 3D treatments.

Including all the physics necessary for realism in a supernova simulation
is a daunting task. The multidimensional simulations mentioned above,
which had simplified neutrino transport, taxed the computational resources
of their time; the same is true of recent simulations involving Boltzmann
transport in spherical symmetry \cite{mezz00,ramp00,lieb00}. Adding
general relativity to the list of desired physics makes things all the
more challenging. While numerical relativity has been successful in 
spherical and axisymmetric cases, ``...in the general three-dimensional
(3D) case which is needed for the simulation of realistic astrophysical
systems, it has not been possible to obtain stable and 
accurate evolutions...,'' and it is argued that the difficulties are
more fundamental than insufficient resolution \cite{alcu00}.
While the difficulties with 3D numerical relativity are beginning to
be overcome \cite{shib99,alcu00b}, the computations are ``prohibitively
slow'' \cite{shib00} when the central gravitating body has a modest compactness
($M/R \lesssim 0.15$). Hence
it will be some time before supernova
simulations with sophisticated microphysics and transport can cover
the full process of collapse, explosion, and wind formation over 
hundreds of milliseconds in a fully relativistic context.

In order to overcome the difficulties associated with
3D general relativistic simulations, and to
save resources for 3D hydrodynamics and accurate neutrino
transport, an approximate treatment of gravity that captures the 
phenomena of dynamical importance in supernovae would be desirable.
The list of new gravitational phenomena introduced by relativity 
includes the nonlinearity of the gravitational field, the inclusion
of all forms of energy and stress as sources, and gravitational waves.
The first two of these effects are of dynamical importance in 
supernovae, while gravitational waves will probably not exert a
strong back-reaction (unless bar or breakup instabilities of some sort
become operative in the core). 
Given the effects we wish to capture, 
what sort of approximation might we try?
A common procedure in the literature has been to assume a spherical
mass distribution in implementing gravity, which allows a relativistic
treatment. However, for rapidly rotating progenitors this introduces
errors of about 10\% during collapse and bounce, with errors rising
at later times as the nascent neutron star cools \cite{frye00}.
A multidimensional approach that captures relativistic phenomena 
could reduce these inaccuracies.
A 3D post-Newtonian approach might be considered, 
but the post-Newtonian limit is known to be 
inadequate for quantitative determinations of neutron star properties.
The case of spherical neutron stars constructed 
with a polytropic equation of state (EOS) with adiabatic index 
$\Gamma=2$ (a common proxy for
more realistic dense matter EOSs) is illustrative:
when the polytropic constant is chosen to make
the maximum mass in a fully relativistic calculation equal to $\sim 1.6\
M_\odot$, the maximum mass determined by first-order post-Newtonian
calculations is $\sim 2.5\ M_\odot$---an error of close to 60\% 
\cite{shin99}.

We conclude that
an approach that both probes the nonlinear nature of gravity more 
deeply, and does so in multidimensions, would be desirable
in the supernova context.
A method that could be incorporated into existing
Newtonian hydrodynamics codes would be even more useful.
We describe such an approximation---which we call ``NewtonPlus''---in \S2, 
displaying
a full set of Einstein equations in order to see where inconsistencies
arise and how serious they are in the supernova environment. Hydrodynamics
equations for the radial direction are presented in \S3 in order to argue
that the approach could be implemented in existing codes. In \S4 
we display results
for rapidly rotating (and therefore highly deformed) stars, which show
that this simple ``NewtonPlus'' approach to gravity is indeed a significant
improvement over the Newtonian limit. Concluding remarks, including 
comments about certain approximate approaches to gravity 
employed in binary neutron star
calculations, are given in \S5. An appendix shows how the static
limit of our
formulation relates to the familiar equations of stellar 
structure in Schwarzschild coordinates.

\section{Einstein Equations in the NewtonPlus Approximation}

As mentioned in \S1, use of the metric of Eq. (\ref{newtonMetric})
in the Einstein equations yields the Newtonian limit, provided the 
gravitational potential $\Phi\ll 1$ and velocities (including microscopic
velocities, large values of which lead to significant stresses) 
are much less than the speed of light. In order to capture the nonlinearity
of gravity, the significance of stresses as gravitational sources, 
and relativistic fluid velocities, we propose the use of the following
metric: 
\begin{equation}
ds^2 = -e^{2\Phi+2\delta} dt^2+e^{-2\Phi}(dr^2 + r^2 d\Omega^2),
\label{newtonPlusMetric}
\end{equation}
where $d\Omega^2\equiv d\theta^2+r^2\sin^2\theta\, d\phi^2$. In comparison
with Eq. (\ref{newtonMetric}), the ``linearized'' metric
functions have been promoted to full exponentials, and a second
metric function, $\delta$,
has been added. Eq. (\ref{newtonPlusMetric}) will
then reduce to the Newtonian case if $\delta\rightarrow 0$; we shall see
that this is in fact the case under conditions
in which the Newtonian limit is valid. This provides a tight conceptual
link with the Newtonian limit. Since it has two independent metric functions,
this ``NewtonPlus'' metric also
provides an exact solution in spherical symmetry. (For the static case,
the Appendix shows the connection between our formulation and the 
familiar equations in Schwarzschild coordinates.)

A convenient formulation of the Einstein equations is the (3+1) formalism,
in which spacetime is foliated into spacelike slices labeled by a time
coordinate \cite{arno62,misn73,smar78,york79}. The metric can be expressed
\begin{equation}
ds^2=-(\alpha^2-\beta_i \beta^i)dt^2 + 2 \beta_i\, dt\, dx^i 
+\gamma_{ij}\,dx^i dx^j,
\label{admMetric}
\end{equation}
where the latin indices run over the three spatial coordinates. As 
originally conceptualized, the four quantities $\alpha$ (the 
lapse function) and  $\beta_i$ (the shift vector) could
be chosen at will as a ``gauge choice'' or as ``coordinate conditions,''
corresponding to 
the invariance of general relativity under coordinate transformations. 
Four constraint equations---the so-called Hamiltonian constraint and
momentum constraints---would be solved to provide self-consistent
data on the initial spacelike slice. The six degrees
of freedom to evolve in time would then be the $\gamma_{ij}$, determined
by equations second order in space and time. Alternatively, one could evolve
equations first order in time for $\gamma_{ij}$ and $K_{ij}$; the latter
is the extrinsic
curvature tensor, which describes the embedding of the spacelike
slices in spacetime. As an alternative to prescribing $\alpha$ and $\beta_i$,
one can, for example, place conditions on $K_{ij}$; then $\alpha$ and $\beta_i$
become quantities for which solutions must be found.

It is apparent that the NewtonPlus metric of 
Eq. (\ref{newtonPlusMetric}) contains only two
of the six degrees of freedom that should be present. This means that
examination of a complete set of Einstein equations should reveal 
inconsistencies; we here explore these and discuss their seriousness in
the supernova environment.

We begin with the Hamiltonian constraint. This yields 
\begin{equation}
\nabla^2\Phi = 4\pi e^{-2\Phi}E + {1\over 2}(\partial\Phi)^2
  - {3\over 2} e^{-4\Phi-2\delta}(\partial_t\Phi)^2.
\label{hamiltonianConstraint}
\end{equation}
In this expression $\nabla^2$ is the usual 3D flat-space
Laplacian. The energy density as 
viewed by an ``Eulerian''
observer (i.e., one whose 4-velocity is orthogonal to the
spacelike slices, having covariant components
$n_\mu = (-\alpha,0,0,0)$, where $\alpha$ is the lapse function)
is denoted by $E\equiv T^{\mu\nu}n_\mu n_\nu$, where $T^{\mu\nu}$ is the
stress-energy tensor. 
For a perfect fluid, $E=\Gamma^2(\rho+p)-p$, where $\Gamma=(1-v^2)^{-1/2}$,
$v$ is the magnitude of the physical fluid velocity as measured by an Eulerian 
observer, and $\rho$ and $p$ are respectively the total energy density
and pressure in the fluid rest frame. We have employed the notation
\begin{equation}
\partial X\,\partial Y\equiv \partial_r X\,\partial_r Y +{1\over r^2}\,
  \partial_\theta X\,\partial_\theta Y +{1\over r^2 \sin^2\theta}\,
  \partial_\phi X\,\partial_\phi Y.
\end{equation}
As expected from the conceptual
link between the Newtonian and NewtonPlus metrics, 
Eq. (\ref{hamiltonianConstraint}) identifies $\Phi$ as a glorified
gravitational potential. In addition to the rest energy, the source
includes internal and kinetic energies and pressure, all boosted by
nonlinear contributions from $\Phi$ itself.

Next we turn to the momentum constraints. These can be summarized by
\begin{equation}
\mbox{\boldmath$ \nabla$}
(e^{-\Phi-\delta}\partial_t \Phi)=-4\pi e^{-\Phi}{\bf s}.
\label{momentumConstraints}
\end{equation}
The three momentum constraint equations are obtained by reading this
as a vector equation in orthonormal spherical coordinates. 
For a perfect fluid,
${\bf s}=\Gamma^2(\rho+p){\bf v}$, where ${\bf v}$ is the physical
velocity measured by an Eulerian observer. Equation 
(\ref{momentumConstraints}) is the first challenge to the consistency
of the use of the NewtonPlus metric:
Since the curl of any gradient vanishes, equation (10) will only
have a solution if the fluid flow is such that
$\mbox{\boldmath$\nabla$}\times(e^{-\Phi}{\bf s})=0$.
This condition is satisfied in spherical symmetry, but it 
typically will not be true when convection is present.
For the use of the NewtonPlus 
metric to be meaningful, there are two possibilities. One is that the
transverse (or solenoidal) portion of $e^{-\Phi}{\bf s}$ is small 
compared with the longitudinal (or irrotational) part, that is,
$\mbox{\boldmath$\nabla$} \times 
(e^{-\Phi}{\bf s})\ll \nabla\cdot(e^{-\Phi}{\bf s})$; this will
obtain if, for example, radial flows dominate convective effects.
In that case, $e^{-\Phi}{\bf s}$ is determined by the Poisson equation
\begin{equation}
\nabla^2(e^{-\Phi-\delta}\partial_t \Phi)=-4\pi \mbox{\boldmath$\nabla$}
\cdot(e^{-\Phi}{\bf s}). \label{dPhidt}
\end{equation}
The second possibility---the one we expect to follow in 
practice---is to neglect $\partial_t \Phi$. 
This is reasonable in supernova environment:
In contrast to the binary
merger problem, one does not have entire stars
moving at relativistic velocities. In the absence of bar or breakup 
instabilities, there is a quasispherical core, with asphericities due
to rotation and convection. Aside from secular changes due to viscosity,
rotation alone can be considered relatively stationary, not contributing 
greatly to explicit time variation of $\Phi$. Core convection may occur,
but probably will involve nonrelativistic velocities. There may be
relativistic radial infall velocities, as seen at late times in
failed explosions \cite{lieb00}, and also  
relativistic
outflow velocities in jets \cite{macf01}; but these situations involve
matter at low density in comparison with the core. Hence in the
various phenomena occuring in the supernova environment there is reason to
suppose that either the densities or velocities are such that
${\bf s}=\Gamma^2(\rho~+~p){\bf v}$ is everywhere small enough for
$\partial_t \Phi$ to be neglected altogether.
Heuristically, ignoring explicit
time derivatives in determining $\Phi$ based on this physical reasoning
is consistent with the expectation that the 
back-reaction due to gravitational
waves will not be dynamically important.

Next, we consider the evolution equations of $\gamma_{ij}$:
\begin{equation}
\partial_t \gamma_{ij} = -2\alpha K_{ij}+\gamma_{jk}D_i\beta^k
+\gamma_{ik}D_j\beta^k,\label{gammaEvolution}
\end{equation}
where $D$ represents a covariant derivative with respect to the 
3-metric $\gamma_{ij}$. For the NewtonPlus metric, the extrinsic
curvature turns out to be 
\begin{equation}
K_{ij}=e^{-\Phi-\delta}\,(\partial_t\Phi)\,
\gamma_{ij}.\label{extrinsicCurvature}
\end{equation}
 Referring back to Eqs. (\ref{newtonPlusMetric})
and (\ref{admMetric}),
it is easy to see that the vanishing shift vector conspires with
the explicit form of the extrinsic curvature tensor to make
Eq. (\ref{gammaEvolution}) an identity leading to no new information.
(This is the case even if we choose not to neglect $\partial_t\Phi$.)

Finally, we come to the evolution equations for $K_{ij}$.
In considering these equations we will take $\partial_t\Phi=0$ in
accordance with the physical reasoning discussed in connection with
the momentum constraints. 
From Eq. (\ref{extrinsicCurvature}), we see that
our neglect of explicit time derivatives of $\Phi$ implies a
vanishing extrinsic curvature tensor. Nevertheless, 
the evolution equations for
$K_{ij}$ provide nontrivial conditions, and will lead
us to an equation for $\delta$.
Various combinations of the Einstein equations could be
employed to give an equation for $\delta$, and we have explored
some of these in numerical computations of stationary rotating
stars. Because of numerical convergence issues
discussed at the end of this section, our experience shows that
convergence is achieved when
the equation for $\delta$ is derived from 
a combination of the individual evolution equations
for $K_{ij}$; the ``lapse equation,'' which is the trace
of the evolution equations for $K_{ij}$; and the Hamiltonian constraint.

The equations for $\delta$ obtained in this manner are as follows.
Combining the evolution equation of $K^r_{\ r}$ with the lapse
equation and the Hamiltonian constraint yields 
\begin{eqnarray}
{{\partial }_r}{{\partial }_r}
   \delta +\frac{1}{r}
    {{\partial }_r}\delta +\frac{1}{2{r^2}\tan \theta }
      {{\partial }_{\theta }}
       \delta +\frac{1}{2{r^2}}
        {{\partial }_{\theta }}
         {{\partial }_{\theta }}
          \delta +\frac{1}{2{r^2}{{\sin^2 \theta }}}
           {{\partial }_{\phi }}
            {{\partial }_{\phi }}
             \delta &=&4\pi
              {{e }^{-2 \Phi }}
                 ( {{{S^{\theta }}}_{\theta }}
               +{{{S^{\phi }}}_{\phi }}
                )-{{({{\partial }_r}\Phi )}^2}\nonumber\\
& &                 -2{{\partial }_r}
                  \Phi {{\partial }_r}
                   \delta -{{({{\partial }_r}\delta )}^2}\nonumber\\
& &                    -\frac{1}{2{r^2}{{\sin [\theta ]}^2}}
                     {{({{\partial }_{\phi }}\delta )}^2}
                     -\frac{1}{2{r^2}}
                     {{({{\partial }_{\theta }}\delta )}^2},
\label{krr}
\end{eqnarray}
while combining the $K^\theta_{\ \theta}$ evolution equation 
with the lapse equation and Hamiltonian constraint yields
\begin{eqnarray}
{{\partial }_r}{{\partial }_r}
   \delta +\frac{1}{r}
    {{\partial }_r}\delta +\frac{1}{{r^2}\tan\theta }
      {{\partial }_{\theta }}
       \delta +\frac{1}{{r^2}{{\sin^2 \theta }}}
        {{\partial }_{\phi }}
         {{\partial }_{\phi }}
          \delta &=&8\pi {{e }^{-2 \Phi }}
              {{{S^{\theta }}}_{\theta }}
            -{{({{\partial }_r}\Phi )}^2}
             +\frac{1}{{r^2}}
              {{({{\partial }_{\theta }}\Phi )}^2}\nonumber \\
& &               -\frac{1}{{r^2}{{\sin^2\theta }}}
                {{({{\partial }_{\phi }}\Phi )}^2}
                 -2{{\partial }_r}
                  \Phi {{\partial }_r}
                   \delta +\frac{2}{{r^2}}
                    {{\partial }_{\theta }}
                     \Phi {{\partial }_{\theta }}
                     \delta\nonumber \\
& & -
                     \frac{2}{{r^2}{{\sin^2\theta }}}
                     {{\partial }_{\phi }}
                     \Phi {{\partial }_{\phi }}
                     \delta -
                     {{({{\partial }_r}\delta )}^2}
                     -\frac{1}{{r^2}{{\sin^2\theta }}}
                     {{({{\partial }_{\phi }}\delta )}^2},
\end{eqnarray}
and similar use of the $K^\phi_{\ \phi}$ evolution equation gives
\begin{eqnarray}
{{\partial }_r}{{\partial }_r}
   \delta +\frac{1}{r}
    {{\partial }_r}\delta +\frac{1}{{r^2}}
      {{\partial }_{\theta }}
       {{\partial }_{\theta }}
        \delta& =&8\pi {{e }^{-2 \Phi }}
            {{{S^{\phi }}}_{\phi }}
          -{{({{\partial }_r}\Phi )}^2}
           -\frac{1}{{r^2}}
            {{({{\partial }_{\theta }}\Phi )}^2}
             +\frac{1}{{r^2}{{\sin^2 \theta }}}
              {{({{\partial }_{\phi }}\Phi )}^2}
               -2{{\partial }_r}
                \Phi {{\partial }_r}
                 \delta  \nonumber \\
& & -\frac{2}{{r^2}}
                  {{\partial }_{\theta }}
                   \Phi {{\partial }_{\theta }}
                    \delta +
                     \frac{2}{{r^2}{{\sin^2\theta }}}
                     {{\partial }_{\phi }}
                     \Phi {{\partial }_{\phi }}
                     \delta -
                     {{({{\partial }_r}\delta )}^2}
                     -\frac{1}{{r^2}}
                     {{({{\partial }_{\theta }}\delta )}^2}.
\label{kpp}
\end{eqnarray}
In these equations, $S_{\mu\nu}\equiv T^{\rho\sigma} h_{\mu\rho} 
h_{\nu\sigma}$, where
the spacelike projection tensor is defined by 
$h_{\mu\nu}\equiv g_{\mu\nu}+n_\mu n_\nu$.
For a perfect fluid, $S^{\theta }_{\ \theta }=(E+p)(v^\theta)^2 + p$,
where $v^\theta$ is the component of the fluid's physical velocity in the
$\theta$ direction as measured by an Eulerian observer,
and $S^{\phi }_{\ \phi }$ is given by a similar expression with $v^\theta$
replaced by $v^\phi$.
Given the fact that we expect inconsistencies in the Einstein equations
due to the reduced number of degrees of freedom that we keep, 
it seems likely that Eqs. (\ref{krr})-(\ref{kpp}) for $\delta$ are 
inconsistent, 
though it is not obvious (to us) how to prove this rigorously.
However, we note that stresses (e.g., pressure) constitute the primary source
terms in the equations determining $\delta$, confirming the expectation
expressed previously, that $\delta$ should vanish as the Newtonian limit
is approached. The observation that $\delta$ will only
be appreciable at the highest densities, where pressure begins to make
a nontrivial contribution in comparison with energy density, suggests
a reasonable path forward. In typical cases it is expected that
the deepest
portion of the core will be roughly spherical, even if rapid rotation
causes an equatorial bulge of lower density material (e.g., Fig. 2 of
Ref. \cite{bona93}). 
If this is the case, neglecting the angular derivatives of $\delta$
is justified, removing many of the apparent inconsistencies in the
three equations for $\delta$. 
The remaining
discrepancies have to do with angular derivatives of $\Phi$ and 
particular components of the stress tensor that appear in 
Eqs. (\ref{krr})-(\ref{kpp}).
As it happens, if one adds (or averages) Eqs. (\ref{krr})-(\ref{kpp}), 
these remaining discrepancies
disappear. Hence the equation we shall use to determine $\delta$ is
\begin{equation}
{{\partial }_r}{{\partial }_r}
   \delta +\frac{1}{r}
    {{\partial }_r}\delta =4\pi
      {{e }^{-2 \Phi }}
         ( {{{S^{\theta }}}_{\theta }}
       +{{{S^{\phi }}}_{\phi }}
        )-{{[{{\partial }_r}(\Phi+\delta) ]}^2}.\label{delta}
\end{equation}
The source on the right-hand side is to be angle-averaged
in solving for $\delta$.
Recalling that $S^{\theta }_{\ \theta }=S^{\phi }_{\ \phi }$
in spherical symmetry, Eq. (\ref{delta}) 
is the equation for $\delta$ obtained in the
spherical case.

It should be straightforward to include the solution of $\Phi$ and,
if desired, $\delta$, in existing multidimensional gravitational 
hydrodynamics codes. 
The solution of $\Phi$ would make use of the Poisson solver normally
used to solve for the Newtonian gravitational potential, the only difference
being that one would have to iterate on 
Eq. (\ref{hamiltonianConstraint}) (with the $\partial_t\Phi$
term dropped) to get a self-consistent $\Phi$.
The simplest approximation would be to simply solve for $\Phi$ in
this manner, and ignore $\delta$ altogether. The next level of
approximation would involve solving Eq. (\ref{delta}) for
$\delta$, but ignoring $\delta$ on the right hand side. Since $\delta$
is already something of a correction, $\delta$ appearing on the right
hand side is essentially a ``correction to the correction.'' If desired,
however, Eq. (\ref{delta}) could be solved as it stands, with
iteration being required. In any case, solving Eq. (\ref{delta}),
while requiring only a couple of angular and radial integrations over
the computational domain, is a bit subtle. This is because 
the Green's function of the operator on the left-hand side 
is a logarithm, which must be delicately handled
in order to get a vanishing boundary condition at infinity \cite{bona93}.
It can be shown that $\delta$ is in fact the same function as that denoted
$\zeta$ in the fully relativistic treatment of axisymmetric stars of Ref.
\cite{card01}, where the proper handling of this subtlety is described
and an integral expression (Green's function expansion) for $\zeta$ 
is given. The first term of this expansion can be adapted as the solution 
to Eq. (\ref{delta}).

\section{Hydrodynamics in the NewtonPlus Approximation}

The utility of the NewtonPlus approximation will be greatly enhanced
if existing multidimensional hydrodynamics codes can be adapted to
its use. As a concrete case we consider the Virginia Hydrodynamics
code (VH-1) \cite{vh1}, an implementation
of the piecewise parabolic method \cite{cole84}. 
This code treats 
a multidimensional problem with operator splitting: Successive sweeps
through the spatial dimensions are taken, with each sweep involving
a Lagrangian step forward in time followed by a remapping of the fluid
variables to a fixed Eulerian grid. 
By way of example, we here
present the equations for a spherically symmetric
calculation in the NewtonPlus metric, which are 
analogous to those solved by VH-1 in its Lagrangian 
time steps in a similar Newtonian calculation. Similar
equations can be derived for use in the angular sweeps.

The basic hydrodynamic equations can be expressed as 
conservation laws which follow from the vanishing divergences
of the baryon flux vector and perfect fluid stress-energy tensor.
It has been known for some time \cite{wils78} that relativistic Eulerian
hydrodynamics can be put into a conservative form amenable to
methods developed for Newtonian hydrodynamics.
In seeking to adapt VH-1 to an approximate relativistic treatment,   
we follow Appendix B of Ref. \cite{lieb99}, which deals with
spherically symmetric relativistic hydrodynamics in Eulerian coordinates. 
It was shown that these Eulerian equations can be cast in a
form that, while not truly ``Lagrangian'' in the sense of being in a
comoving frame, is nevetheless similar to the Newtonian Lagrangian
equations. 

Consider
a vector of state variables,
\begin{equation}
{\bf u}\equiv (D,S,E).
\end{equation}
The components of this vector are defined by
\begin{eqnarray}
D\equiv \Gamma \rho_B, \\
S\equiv \Gamma^2(\rho+p)v,\\
E\equiv \Gamma^2(\rho+p)-p,
\end{eqnarray}
where $\Gamma\equiv(1-v^2)^{1/2}$ 
is the boost factor between the fluid rest frame
and the fixed ``Eulerian'' frame; $\rho_B$, $\rho$, and $p$ 
are respectively the 
baryon rest mass density, total energy density, and pressure
in the fluid rest frame;
and $v$ is the fluid physical radial velocity as measured by an Eulerian
observer. This state vector is governed by the hydrodynamics equations
in the metric of Eq.~(\ref{newtonPlusMetric}),
\begin{equation}
\partial_t{\bf u}+{e^{3\Phi}\over r^2}\,\partial_r(r^2 e^{\delta-\Phi}{\bf f})
= \mbox{\boldmath$\sigma$},
\label{hydro1}
\end{equation}
where the fluxes are
\begin{equation}
{\bf f}=(D v,S v+p, S),
\end{equation}
and the sources are
\begin{eqnarray}
\sigma_D&=&3D\, \partial_t\Phi,\\
\sigma_S&=&-e^{\delta+2\Phi}(E+p)\,\partial_r(\Phi+\delta) + 
e^{\delta+2\Phi} p\left[{2\over r}+\partial_r(\delta-\Phi)\right]
+4 S\, \partial_t\Phi,\\
\sigma_E&=&-e^{\delta+2\Phi} S\,\partial_r(\Phi+\delta)+
  (3p + S v+3 E)\,\partial_t\Phi.
\end{eqnarray}
It is convenient to separate out the terms involving the advection of
the state variables; to this end, Eq. (\ref{hydro1}) can be
rewritten as
\begin{equation}
\partial_t{\bf u}+{e^{3\Phi}\over r^2}\,\partial_r(r^2 e^{\delta-\Phi}{\bf u}v)
+{e^{3\Phi}\over r^2}\,\partial_r(r^2 e^{\delta-\Phi}{\bf k})
= \mbox{\boldmath$\sigma$},
\label{hydro2}
\end{equation}
where the vector
\begin{equation}
{\bf k}=(0,p, pv)
\end{equation}
has been defined.

Next we consider how the conservation laws of Eq. (\ref{hydro2})
can be treated by a hydrodynamics code like VH-1, which
performs Lagrangian hydrodynamic time steps followed by a ``remap''
to an Eulerian grid.
We begin with baryon number conservation. 
The baryon flux vector is $A^\mu=\rho_B u^\mu$, where $u^\mu$ is the
fluid 4-velocity.
The number
of baryons in a proper three-volume described by the 1-form $\Delta\Sigma_\mu$
with edges $\Delta r$, $\Delta \theta$, and $\Delta\phi$ is
\begin{equation}
{\rm constant}=A^\mu\,\Delta\Sigma_\mu=(e^{-3\Phi}r^2\sin\theta\,\Delta r\,
\Delta\theta\,\Delta\phi)\Gamma \rho_B.
\label{baryonConservation}
\end{equation}
In VH-1, 
the mass density is updated during a Lagrangian step as follows. 
From the time averaged
fluid velocities at the zone edges obtained from the solution of the
Riemann problem, the new positions of the zone edges 
are computed. From the new zone edge positions, the new
zone volume is computed, and the new zone density is given by
\begin{equation}
{\rm (new\ density)=(old\ density){(old\ zone\ volume)
\over(new\ zone\ volume)}}.
\end{equation}
Clearly, Eq. (\ref{baryonConservation}) can be used in precisely
the same way. 
There is a factor of $\Gamma$ that must be dealt with, but
presumably a relativistic Riemann solver \cite{mart96}
can be used if allowance for
relativistic velocities is desired.

In VH-1 the updates of the velocity $v$ and specific internal energy $e$ make
use of the (Newtonian) Lagrangian equations
\begin{eqnarray}
\partial_t v+r^2\,\partial_m p& =& g, \label{velocity}\\
\partial_t e + \partial_m(r^2 p v)&=&v\, g.\label{energy}
\end{eqnarray}
Here the mass coordinate is defined 
by
\begin{equation}
m=\int_0^r \rho_B\, r^2\,dr,
\end{equation}
and $g$ is a force---gravity, for instance. 

In the NewtonPlus approximation a similar set of equations can be obtained.
We transform from the variables $t,r$ to $\bar t,m$, where
\begin{eqnarray}
\bar t&=& t,\label{coordinateTransform1}\\
m &=& \int_0^r\Gamma\rho_B\, e^{-3\Phi}r^2\,dr=\int_0^r D e^{-3\Phi} r^2\,dr.
\label{coordinateTransform2}
\end{eqnarray}
Even though $\bar t=t$, we still write derivatives as $\partial_{\bar t}$
or $\partial_t$ to indicate whether $m$ or $r$ is being held constant
respectively. We now use this change of coordinates to bring equation
(\ref{hydro2}) into a useful form. The first component of equation
(\ref{hydro2}) is
\begin{equation}
\partial_t D+{e^{3\Phi}\over r^2}\partial_r(r^2 e^{\delta-\Phi}D v)=
3D\, \partial_t\Phi.
\label{baryonConservation2}
\end{equation}
Subtracting ${\bf u}/D$ times Eq. (\ref{baryonConservation2}) from 
Eq. (\ref{hydro2}) yields
\begin{equation}
D\,\partial_t\left({\bf u}\over D\right)+
D e^{\delta+2\Phi}v\,\partial_r\left({\bf u}\over D\right)
+{e^{3\Phi}\over r^2}\,\partial_r(r^2 e^{\delta-\Phi}{\bf k})
= \mbox{\boldmath$\sigma$}-3{\bf u}\, \partial_t\Phi.
\label{hydro3}
\end{equation}
Equations (\ref{coordinateTransform1}-\ref{baryonConservation2}) 
can be used to show that 
\begin{eqnarray}
\partial_t&=&\partial_{\bar t} - e^{\delta-\Phi}r^2 v D\, \partial_m,\\
\partial_r&=&r^2 e^{-3\Phi}D\, \partial_m,\label{partialm}
\end{eqnarray}
so that Eq. (\ref{hydro3}) can be expressed as
\begin{equation}
\partial_{\bar t}\left({\bf u}\over D\right)+
\partial_m(r^2 e^{\delta-\Phi}{\bf k})
= {\mbox{\boldmath$\sigma$}\over D}-3{\bf u}\,\partial_t\Phi.
\label{hydro4}
\end{equation}
The second and third components of this equation are
\begin{eqnarray}
\partial_{\bar t}\left(S\over D\right)+
r^2 e^{\delta-\Phi}\,\partial_m p
&=& g+{S\over D}\,\partial_{t}\Phi,\label{velocity2}\\
\partial_{\bar t}\left(E\over D\right)+
\partial_m(r^2 e^{\delta-\Phi}p v)
&=&v\, g +{3p+S v\over D}\,\partial_{t}\Phi, \label{energy2}
\end{eqnarray}
where the ``gravitational force'' $g$ is defined by
\begin{equation}
g=-e^{\delta+2\Phi}{(E+p)\over D}\,\partial_r(\Phi+\delta)=
-e^{\delta-\Phi}{(E+p)}\,\partial_m(\Phi+\delta).\label{gravity}
\end{equation}
In keeping with our practice of dropping explicit time derivatives of
$\Phi$ (at constant $r$), 
the last two terms of Eqs. (\ref{velocity2}) and
(\ref{energy2}) can be neglected. Furthermore,
it is convenient\footnote{Computationally, this prevents finite
differences in the internal
energy from being swamped by finite differences of the (rest+internal) energy.}
 to define a
specific internal energy $U$ by
$U\equiv E-D$, so that
 on the left hand
side one can write $\partial_{\bar t}\left(E/ D\right)=
\partial_{\bar t}\left[(U/ D)+1\right]=\partial_{\bar t}\left(U/ D\right)$. 

The resulting equations,
\begin{eqnarray}
\partial_{\bar t}\left(S\over D\right)+
r^2 e^{\delta-\Phi}\,\partial_m p
&=& g,\label{velocity3}\\
\partial_{\bar t}\left(U\over D\right)+
\partial_m(r^2 e^{\delta-\Phi}p v)
&=&v\, g, \label{energy3}
\end{eqnarray}
are of
the same form as Eqs. (\ref{velocity}) and (\ref{energy}),
allowing similar computational methods to be employed in 
their solution. 
It may seem surprising that the equations look so similar
in the NewtonPlus and Newtonian formulations. As mentioned
previously, it has
been known for some time \cite{wils78} that relativistic 
conservation laws can be cast
in a ``conservative'' form similar to the Eulerian Newtonian equations. 
Because we have taken care to define a mass coordinate based on
the proper relativistic 3-volume element, and chosen a metric with 
a close connection to the one giving the Newtonian limit, 
we have also been able to find equations quite close to the
Newtonian Lagrangian formulation.

\section{Testing NewtonPlus Gravity with Rapidly Rotating Stars}

In this section we present calculations of neutron stars undergoing
rapid uniform rotation in order to assess the strengths and weaknesses
of the NewtonPlus approximation to relativistic gravity. Our
models were computed with a code described in Ref. \cite{card01}, which
was written to compute the structure of relativistic axisymmetric stars.
We have modified the code to include the ability to perform computations
in the Newtonian and various NewtonPlus limits: with vanishing metric
function $\delta$, with ``linearized'' $\delta$ (i.e. ignoring $\delta$ on
the right hand side of Eq. (\ref{delta})), and ``full'' $\delta$
(solving Eq. (\ref{delta}) as written). All of the NewtonPlus limits
solve a two-dimensional (and stationary) version of the
nonlinear Poisson-type Eq. (\ref{hamiltonianConstraint}) for the
glorified ``gravitational potential'' $\Phi$. For the results presented here,
the high-density portion of
the equation of state (EOS) is taken from Ref. \cite{pcl95}, and is based on a
field-theoretic description of cold dense matter. 
We also performed calculations
with a polytropic EOS of adiabatic index 2, and found qualitatively similar
results.

Panel (a) of Fig. \ref{pcl} 
shows mass vs. radius curves for spherical stars. 
The gravitational mass, a measure of total mass-energy, is defined 
for asymtotically flat spacetimes by
(see e.g. Ref. \cite{bona93})
\begin{equation}
M=\int({T^i}_i-{T^t}_t)\sqrt{-g}\,dx^1 dx^2 dx^3,
\end{equation}
where here $g$ is the determinant of the metric and $x^i$ are
spatial coordinates. For the NewtonPlus metric,
\begin{equation}
M=\int e^{\delta-2\Phi}(E+{S^i}_i)r^2\sin\theta\,dr\, d\theta\, d\phi.
\end{equation}
From Eq. (\ref{newtonPlusMetric}), it is evident that the proper
equatorial radius is 
\begin{equation}
R=\exp[{-\Phi(r=r_{\rm surface},\theta=\pi/2)}]\, r_e,
\end{equation}
where $r_e$ is the coordinate radius of the equatorial
surface.
Panel (a) shows that while the Newtonian limit exhibits no maximum
mass with this EOS,\footnote{No turnover in the mass vs. radius curve
appears in the Newtonian limit, 
up to the high-density boundary of the tabulated EOS. 
A configuration with
central baryon mass density (total energy density) of 
$3.07\times 10^{15}$ g cm$^{-3}$ ($4.65\times 10^{15}$ g cm$^{-3}$)
has a gravitational mass of 15.6 $M_\odot$ and radius 18.9 km in 
the Newtonian limit, while the relativistic configuration with this 
central density has a gravitational mass of 1.68 $M_\odot$
and a radius of 9.40 km. We remind the reader that the Chandrasekhar mass
phenomenon is a property of stars built on a polytropic equation of state
with adiabatic index equal to 4/3, and that ``realistic'' nuclear equations
of state will not generally exhibit this behavior. Instead, the upper mass
limit of neutron stars derives from the the general relativistic instability
indicated by the turning point in the mass vs. radius curve.} 
the NewtonPlus approximation does yield a maximum mass. 
Even with vanishing $\delta$, the approximation captures this consequence
of nonlinear gravity. The ``linear $\delta$'' approximation follows the exact
relativistic curve until the most dense configurations are reached,
where the approximation ``overshoots'' the true mass (i.e. masses are
too large by a significant factor when $\delta$ is neglected, 
and a bit too small when the ``linear $\delta$'' approximation is employed).
Since
pressure is the main source for $\delta$ (see equation (\ref{delta})),
the large pressures associated with such high densities raise $\delta$ to 
large enough values that it cannot be neglected on the right-hand side of
equation (\ref{delta}). 
As expected, 
the ``full $\delta$'' 
approximation is indistinguishable from the relativistic results
in spherical symmetry, where only two metric functions are needed to
describe the spacetime exactly. The behavior of the ``vanishing $\delta$''
and ``linear $\delta$'' approximations in comparison with the exact result
is given further explanation in the Appendix.

Panels (b)-(f) of Fig. \ref{pcl} show various physical parameters of rapidly
rotating configurations. In order to test the NewtonPlus approximation
in a nonspherical setting, we ask the question: Given a definite number
of baryons rotating at a given uniform angular velocity $\Omega$, 
what do the various treatments of gravity do with those baryons? 
(The fact that baryon number is a conserved quantity makes this an
obvious way to compare different treatments of gravity.)
To 
answer this question we have computed, for each treatment of gravity, 
a constant baryon mass sequence
beginning at zero rotation (marked by squares) and ending at the
mass shedding limit (marked by stars). The value of baryon mass chosen,
1.8 $M_\odot$, is close to the maximum baryon mass of 1.95 $M_\odot$ 
for the equation of state we employed. The baryon mass is defined by
(e.g., Ref. \cite{bona93})
\begin{eqnarray}
M_B&=&\int(-n_\mu A^\mu)\sqrt{h}\,dx^1 dx^2 dx^3,\\
&=&\int e^{-3\Phi}\,\Gamma \rho_B\, r^2\sin\theta\,dr\, d\theta\, d\phi,
\end{eqnarray}
where $n_\mu$ is defined after Eq. (\ref{newtonPlusMetric}),
$A^\mu$ is defined before Eq. (\ref{baryonConservation}), and
$h$ is the determinant of $h_{\mu\nu}=g_{\mu\nu}+n_\mu n_\nu$.
The second line (here and in the equations below) is specialized
to the NewtonPlus metric.
The quantities plotted as a function of the (uniform)
stellar angular velocity in panels (b)-(f) are
(b) gravitational mass; (c) equatorial radius; (d) angular momentum, given by
(e.g., Ref. \cite{bona93})
\begin{eqnarray}
J&=&\int{T^t}_\phi\sqrt{-g}\,dx^1 dx^2 dx^2,\\
&=&\int e^{-\delta-6\Phi}(E+p)\,\Omega\, r^4\sin^3\theta\,dr\, d\theta\, d\phi;
\label{angularMomentum}
\end{eqnarray}
(e) the eccentricity, defined here as $1-r_p/r_e$, where $r_p$ and $r_e$ are
respectively the polar and equatorial coordinate radii; and (f) the
linear equatorial velocity, given by (e.g. Ref. \cite{bona93})
\begin{eqnarray}
U&=&{1\over\Gamma}(e_\phi)_\mu u^\mu \\
&=& e^{-\delta-2\Phi}r\sin\theta\;\Omega,\label{ueq}
\end{eqnarray}
where $(e_\phi)^\mu$ is the basis vector corresponding to the coordinate
$\phi$. 

In panels (b)-(f), the efficacy of the NewtonPlus approximations
can be judged by choosing a value of angular velocity and seeing how close
the approximate quantities come to the fully relativistic value.
While the ``full $\delta$'' approximation is indistinguishable from full
relativity in the spherical case, the two curves representing these
treatments deviate from one another with increasing angular velocity.
In panels (b) and (c), the ``overshoot'' of the ``linear $\delta$''
approximation, discussed in connection with panel (a), is again visible
at low angular velocities. At the highest angular velocities, the
``linear $\delta$'' approximation actually appears to give slightly better
results than the ``full $\delta$'' calculations, but it must be remembered
that this is a lingering result of the ``overshoot'' at low angular
velocities. 
As expected, the angular velocities at mass shedding
of the NewtonPlus approximations are closer to the relativistic values
than the Newtonian case. The NewtonPlus treatments are quite successful at 
approximating the gravitational mass, radius, and eccentricity, while 
the success of the results for angular momentum and equatorial velocity
is more modest. 

The relative success of the NewtonPlus approximation in determining these
observables can be understood by examining expressions for these
quantities in the fully relativistic case.
Axisymmetric, stationary spacetimes can be described by four metric functions;
the metric can be expressed \cite{bona93,card01}
\begin{equation}
g_{\alpha\beta}dx^\alpha dx^\beta = -e^{2\nu}dt^2 + e^{-2\nu}G^2
	r^2 \sin^2 \theta (d\phi - N^\phi dt)^2
	+ e^{2(\zeta-\nu)}(dr^2 + r^2 d\theta^2).\label{relativisticMetric}
\end{equation} 
In comparison, the NewtonPlus metric is conformally flat (equivalent
to requiring $G=e^{\zeta}$ in Eq. (\ref{relativisticMetric})),
and lacks the shift vector component $N^\phi$. In the
relativistic case the angular momentum and equatorial velocity are
given by
\begin{eqnarray}
J&=&\int e^{2\zeta-6\nu}G^3(E+p)(\Omega-N^\phi)r^4\sin^3\theta\,
  dr d\theta d\phi,\\
U&=&e^{-2\nu}G r\sin\theta(\Omega-N^\phi).
\end{eqnarray}
In comparison with Eqs. (\ref{angularMomentum}) and (\ref{ueq}),
an important difference is that $\Omega$ is replaced by $\Omega-N^\phi$.
Since values of $N^\phi$ can be a significant fraction of 
$\Omega$
in extreme configurations (e.g., Refs. \cite{bona93,bona94}),
the neglect of this shift vector component in the NewtonPlus approximations
probably accounts for most of the difference in $J$ and $U$ with 
the relativistic case.\footnote{Since the lapse function and shift
vector were quantities to be chosen at will in the (3+1) 
formalism as originally
conceptualized, at first glance it may seem strange that the 
shift vector component $N^\phi$ should play a significant role in
the determination of physical quantities. 
However, other coordinate choices
have been made in arriving at
Eq. (\ref{relativisticMetric}), making the lapse and shift vector
component quantities with physical content for which solutions must
be found. Specifically, two of the four degrees of coordinate
freedom have been used to choose the time coordinate $t$ to
label the spacelike hypersurfaces which are invariant under time translations,
and the azimuthal coordinate
$\phi$ to measure the angle about the axis of symmetry.
Once these two choices have been made, the metric components
$g_{tr}$, $g_{t\theta}$, $g_{\phi r}$, and $g_{\phi\theta}$ vanish
(see Ref. \cite{bard70}, which draws on \cite{cart69}). The remaining
two degrees of coordinate freedom have been used to set $g_{r\theta}=0$
and $g_{\theta\theta}=r^2 g_{rr}$.} 
The gravitational and baryon mass in the relativistic
case are given by
\begin{eqnarray}
M&=&\int e^{2(\zeta-\nu)}\,G(E+{S^i}_i+2e^{-\nu}N^\phi J_\phi)
r^2\sin\theta\,dr\, d\theta\, d\phi,\label{relativisticM}\\
M_B&=&\int e^{2\zeta-3\nu}\,G\,\Gamma \rho_B\, 
r^2\sin\theta\,dr\, d\theta\, d\phi,\label{relativisticMb}
\end{eqnarray}
where $J_\phi=(E+p)e^{-\nu}\,G\,r\sin\theta\,U$, and the proper
equatorial radius is 
\begin{equation}
R=\left(e^{-\nu}G|_{r=r_{\rm surface},\theta=\pi/2}\right)
\, r_e.\label{relativisticR}
\end{equation}
In Eqs. (\ref{relativisticMb}) and (\ref{relativisticR}), $N^\phi$ does
not appear at all, and it plays a minor role in Eq. (\ref{relativisticM}).
This, together with the fact that the approximation of conformal flatness
is known to work well for rapidly rotating stars \cite{cook96}, shows
why the NewtonPlus approximation gives good results for the gravitional
mass, baryon mass, and equatorial radius.

\section{Conclusion}

Accurate neutrino transport, 3D hydrodynamics, and
relativity are all essential for realistic supernova simulations.
Given the constraints of current hardware, these cannot all be
treated simultaneously with the detail they deserve. We have therefore
presented an approximation to full relativity 
(or set of related approximations) that captures important
relativistic effects in the quasispherical supernova environment:
nonlinearity creating a deeper potential well, and pressure and
internal energy density being 
nontrivial sources of gravitation.

This ``NewtonPlus'' approach to gravity has a tight conceptual link
with the Newtonian limit that yields certain advantageous features. 
The gravitational portion of 
multidimensional Newtonian calculations involves
only the solution of the Poisson equation for the gravitational potential
$\Phi$, and taking its gradient to find the gravitational force. 
The basic idea of our NewtonPlus approach is to promote the Newtonian metric 
functions---which are linear in  $\Phi$---to full exponentials. We also add
a second metric function, $\delta$, whose main source is pressure; hence this
metric function vanishes in the Newtonian limit.    
The Einstein
equations yield a nonlinear Poisson-type equation for a (now glorified)
``gravitational potential,'' whose solution in 3D can be obtained
in a manner similar 
to what would be done in the Newtonian limit. The inconsistencies in 
the Einstein equations arising from the reduced number of degrees of freedom
turn out to be relegated to the subdominant metric function 
$\delta$; they can be 
removed by ignoring angular variations in $\delta$. This is expected
to be successful in the
supernova context because the region where $\delta$ makes the greatest
difference is where pressure is significant in comparison with
rest mass density. Normally, this is the deepest portion of the 
collapsed core, which is roughly 
spherical even when the outer layers bulge at the equator 
due to rapid rotation.\footnote{Ultra-strong magnetic fields
\cite{bocq95} or differential rotation (e.g., Ref. \cite{baum00})
can give rise to off-center density maxima, making the spherical
correction for stresses via the metric function $\delta$ less useful.
The NewtonPlus approximation with $\delta=0$ could still be employed
in such (probably exceptional) cases, however.}
This strategy---allowing the main contribution to the gravitational field
to be multidimensional and nonlinear, while allowing a spherical correction
for the contribution of stresses---reproduces fairly well 
many of the physical characteristics of rapidly rotating 
(stationary) relativistic stars.
In future dynamic
calculations, we intend to ignore explicit time derivatives of $\Phi$
that appear in the Einstein equations, and we have given specific arguments
for the validity of this approximation in the supernova environment.
(This can be tested in the future by comparing the results of 
spherically symmetric collapse simulations using NewtonPlus gravity
with similar fully relativistic computations.)  
Importantly, the hydrodynamics equations in spherical symmetry
obtained from the NewtonPlus metric
are of the same form as those used in a popular Newtonian hydrodynamics
algorithm, providing the expectation that existing Newtonian codes might
be adapted fairly easily to the NewtonPlus approach.

Finally, we make a few comments on approximations to full relativity
employed by Mathews and Wilson \cite{math00} 
and by Shibata, Baumgarte, and Shapiro \cite{shib98}
in binary neutron star
merger calculations. While to our knowledge these
approaches have not been implemented in supernova simulations, they
conceivably could be; hence we offer a few comments by way of comparison.
These approaches involve the use of a conformally flat metric, the 
imposition of a traceless extrinsic curvature tensor, and
the
neglect of explicit time derivatives of gravitational 
variables. Some arguments in favor of these approximations are
given in 
Ref. \cite{math98}.\footnote{Mathews and Wilson's 
initial calculations produced
the controversial result that the individual stars collapsed to black
holes prior to merger, and it was suspected that this might be caused
by their approximations rather than a physical effect.
However, an error in their equations was pointed out in Ref.
\cite{flan98}, and the discrepancies with previous expectations and the 
results of other groups' calculations have been greatly mitigated in their
most recent results \cite{math00}.} These approximations are similar to
ours in that both retain a reduced number of degrees of freedom
(they keep four, while we keep two), both assume conformal flatness,
and both ignore explicit time derivatives of gravitational variables.
An important difference between Refs. \cite{math00} and \cite{shib98}
is that the latter neglect certain nonlinear gravitational terms that
do not arise in the limit of spherical symmetry.
We would argue that an advantage of our approach is the conceptual link
with the Newtonian limit. In the quasispherical supernova context this
makes for only one dominant metric function---the glorified ``gravitational
potential''---while in the other approaches both the conformal factor
and the lapse function are significant and must be solved for in 
multidimensions. They retain the entire shift vector; we have suggested
dropping it altogether, though our work shows that this causes errors in
the angular momentum and equatorial velocities. In the context of
assumed conformal flatness, Wilson and Mathews also found that 
proper treatment of the shift vector was critical to accurate treatment
of the angular momentum  \cite{math00}. 
While angular momentum is perhaps less critical
in supernovae than in binary mergers, it is known that the assumption
of conformal flatness alone (while retaining an azimuthal shift vector
component) yields accurate results for rapidly rotating stars
\cite{cook96}. This suggests a future extension to the work presented here
that would improve results for angular momentum and equatorial velocity.
The
azimuthal shift vector component $N^\phi$ obeys an elliptic equation, whose
solution can be written as a
Green's function expansion (see Ref. \cite{card01} for an
explicit formula). Like the
solution of Eq. (\ref{delta}) for $\delta$, obtaining the leading
term of $N^\phi$ would involve only a pair of angular and 
radial integrations over the
spacetime, but it would likely improve results for angular momentum
and equatorial velocity significantly.

\acknowledgements{We thank O. E. B. Messer 
and W. R. Hix for 
discussions. C.Y.C. is supported by a DoE PECASE award, and A.M.
is supported at Oak Ridge National Laboratory, managed by UT-Battelle,
LLC, for the U.S. Department of Energy under contract DE-AC05-00OR22725.}

\appendix

\section*{}

This appendix shows the relationship between the NewtonPlus metric
in the static limit and the (interior)
Schwarzschild metric. We point out 
that the equation of hydrostatic equilibrium 
derived with the NewtonPlus metric is
equivalent to the familiar Tolman-Oppenheimer-Volkov (TOV) equation,
as expected if the NewtonPlus metric provides an exact solution in
spherical symmetry.  
The connection to the Schwarzschild metric also allows greater insight
into the behavior
of the ``vanishing $\delta$'' and ``linear $\delta$'' approximations in
the static limit. 

\def\rbar{\overline r}
\def\phib{\overline\Phi}

The Schwarzschild line element is given by 
\begin{equation}
ds^2=-e^{2\phib}dt^2+\left(1-{2m\over \rbar}\right)^{-1} d\rbar^2 +\rbar^2
d\Omega^2,
\end{equation}
where $\phib$ and $m$ are functions of $\rbar$. Comparing with a static
version of Eq. (\ref{newtonPlusMetric}) in which $\Phi$ and $\delta$
are only functions of $r$, we find that equivalence between the two
metrics requires
\begin{eqnarray}
\rbar&=&e^{-\Phi} r,\label{rbar}\\
\phib&=&\Phi+\delta,\\
\left(1-{2m\over \rbar}\right)^{-1}&=&\left(1+\rbar{d\Phi\over d\rbar}\right)^2
=\left(1+r{d\Phi\over dr}\right)^{-2}\label{grr}.
\end{eqnarray}

We now describe how equations derived from the NewtonPlus metric are equivalent
to the TOV equation,
\begin{equation}
{dp\over d\rbar}=-{\rho+p\over\rbar}\left(1-{2m\over \rbar}\right)^{-1}\left(
{m\over \rbar}+4\pi\rbar^2 p\right).
\end{equation}
The $rr$ component of the Einstein equations in the static NewtonPlus metric is
\begin{equation}
{2\over r}{d\delta\over dr}-2{d\delta\over dr}{d\Phi\over dr}
-\left({d\Phi\over dr}\right)^2=8\pi e^{-2\Phi}p.\label{einsteinrr}
\end{equation}
Using Eqs. (\ref{rbar}), (\ref{grr}), and (\ref{einsteinrr}), it is
a straightforward exercise to show that the TOV equation becomes
\begin{equation}
{dp\over dr}=-(\rho+p)\left({d\Phi\over dr}+{d\delta\over dr}\right),
\end{equation}
which is the static version of Eq. (\ref{velocity2}) (see also Eqs.
(\ref{partialm}) and (\ref{gravity})).

It is well known that $m(\rbar)$ evaluated at the stellar surface is
equal to the gravitational mass. Using Eqs. (\ref{grr}) and 
(\ref{einsteinrr}), we find
\begin{eqnarray}
{m\over\rbar}&=&r{d\Phi\over dr}-{r^2\over 2}\left({d\Phi\over dr}\right)^2
\label{m1}\\
&=&r{d\Phi\over dr}-{r^2\over 2}\left[\left(
{2\over r}{d\delta\over dr}-2{d\delta\over dr}{d\Phi\over dr}
\right)-8\pi e^{-2\Phi}p\right]\label{m2}.
\end{eqnarray} 
By evaluating these expressions at the stellar surface, we gain
insight into the behavior of the ``vanishing $\delta$'' and
``linear $\delta$'' limits. In the exact solution, the second term
in Eqs. (\ref{m1}) and (\ref{m2}) is negative. In the ``vanishing $\delta$''
approximation, this term becomes positive, explaining why the gravitational
masses obtained in this limit are too large. Because the exact second term
is negative, the quantity in parentheses in Eq. (\ref{m2}) must be positive
and greater in magnitude than the pressure term. Also, since $\delta$ is
subdominant, Eq. (\ref{einsteinrr}) shows that $d\delta/dr$ is positive.
The ``linear $\delta$'' approximation 
involves neglecting terms containing $\delta$ that are
nonlinear in metric
functions, which here involves neglecting
$2(d\delta/dr)(d\Phi/dr)$. This makes the quantity in parentheses 
in Eq. (\ref{m2}) more positive, which makes the overall second term
too negative. This explains the ``overshoot'' (gravitational masses
too small) seen this limit.


\begin{figure}
\caption{Panel (a): Mass vs. radius curves for spherical configurations
  computed with various treatments of gravity. Panels (b)-(f): Various
  physical quantities characterizing uniformly rotating configurations,
  plotted as a function of angular velocity. Each curve represents
  a constant baryon mass sequence computed with the treatments of gravity
  labeled in panel (a).}
\label{pcl}
\end{figure}

\epsfig{file=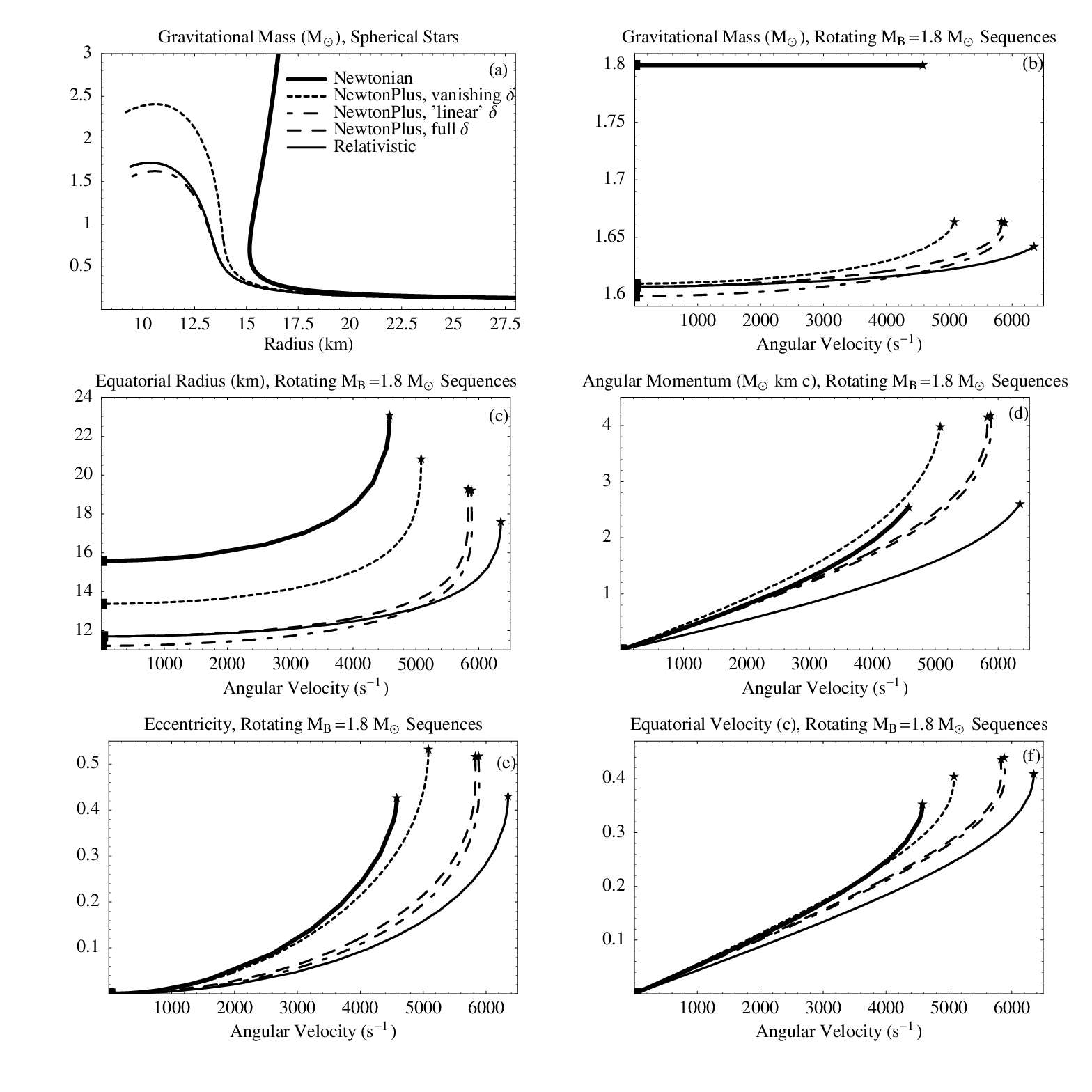,width=6.0in}

\end{document}